\documentclass[rmp,12pt,a4paper]{revtex4}
 \usepackage{graphicx}
 \usepackage{epstopdf}
 \usepackage{amsmath,amssymb}
 \usepackage[colorinlistoftodos]{todonotes}
 \usepackage{natbib}
 
   \usepackage{fancyhdr}
 \fancyfootoffset{0in}
\begin{document}
 
\title{Asking gender questions: Results from a survey of gender and question asking among UK Astronomers at NAM2014}

\fancyhead{}
 \fancyfoot{}
 \cfoot{\thepage}
 \renewcommand{\headrulewidth}{0pt}
 

\maketitle

{\bf Jonathan Pritchard, Karen Masters, James Allen, Filippo Contenta, Leo Huckvale, Stephen Wilkins and Alice Zocchi report on a survey of the gender of astronomers attending and asking questions at the UK National Astronomy Meeting held in Portsmouth in June 2014.}

Inspired by a recent report on the gender balance of astronomers asking questions at the 223rd Meeting of the American Astronomical Society (AAS), held in Washington, DC , January 2014 \citep{davenport2014} we decided to repeat the experiment at the most recent National Astronomy Meeting (NAM). 

The gender balance of both speakers and session chairs at NAM (31\% and 29\% women respectively) closely matched that of attendees (28\% female) however we find that women were under-represented among question askers (which were made up of just 18\% women). Women were especially under-represented in asking the first question (with just 14\% of first questions asked by women), but when the Q\&A session reached four or more questions, women and men were observed to ask roughly equal numbers of questions. 
We found a small increase in the fraction of questions from women in sessions where the chair was also female (but this had no statistical significance). We find that on average $2.2\pm0.1$ questions were asked per talk, and observed no detectable difference in the number of questions asked of female and male speakers, but found that female chairs solicited slightly fewer questions on average than male chairs. 

These results have some similarities and some subtle differences to those reported by \citet{davenport2014} for the AAS. They also found that the gender balance of speakers and chairs closely matched that of attendees (all roughly 35\% women), and that women were under-represented among question askers (24\% women). However, \citet{davenport2014} found a significant effect that when a session chair was female, women asked more questions, and also found that women speakers were asked more questions (with an average of $N=3.2\pm0.2$ questions per female speaker) than their male counterparts (where the average was $N=2.6\pm0.1$). We note that Davenport et al. also report that more questions are asked on average at AAS talks than we found to be asked at NAM.

\section{Survey Method}

The National Astronomy Meeting (NAM) is the largest UK astronomical conference and as such provides a good opportunity to obtain representative statistics on the demographics and dynamics of the UK astronomy and geophysics community and their behaviour at a professional astronomical conference. The NAM2014 participant list contained 624 names from universities and organisations primarily in the UK. Using an open source Python module\footnote{https://pypi.python.org/pypi/SexMachine/}, supplemented by manual classification, we obtained an estimate of the gender breakdown of NAM2014 attendees based on the first names, which results in 72\% male (452), 28\% female (172). 

The scientific programme at NAM was organised into 8 plenary talks, a town hall meeting, and approximately 63 parallel sessions which together had 363 talks (for a review of the meeting see \citealt{bowler2014}). We used a modified version of the web form created by  \citet{davenport2014} to collect data on the gender of speakers, chairs and question askers in as many NAM talks as possible. The form asked for a talk ID (a concatenation of NAM session ID plus talk order was suggested), gender of the speaker, gender of the chair, and then for the input of a string representing the gender of people asking questions (e.g. `FFM'  would be entered if three questions were asked, the first two by women, the last by a man). We added a free form comment box (following the recommendation in  Davenport et al.), but no comments were submitted.  Data collection was volunteer-led, requested via an email to participants from the NAM Local Organizing Committee before the conference began, and a daily Twitter campaign on the \texttt{\#rasnam2014} hashtag. By the end of NAM this form had collected 595 separate submissions corresponding to 263 unique talks, which represented more than 70\% of the scientific content at NAM. On average we collected 2.3 responses per talk. The completeness of this dataset is significantly higher than that analysed by  \citet{davenport2014}, who collected 300 responses comprising data on 225 talks, or about 26\% of talks at the AAS meeting. 

During the NAM Hack Day \citep{simpson2014} we began analysing the data. The raw data required extensive cleaning to correct for non-unique talk IDs, caused both by different naming of sessions and different numbering of talks within a session. Since we collected multiple answers per talk we were also faced with examples where they did not agree. Where a talk had more than one response, and there was no consistent majority answer, we chose to retain the longer question strings under the assumption that shorter ones represented abbreviated or premature submissions of the form. We note that differences could also be due to ambiguity of what response is appropriate, for example if the same questioner asks two questions at the same time, or if there was a back and forth between speaker and questioner. This was a factor for just a small fraction of entries so we do not expect it to influence our overall conclusions.

\section{The Gender Balance of NAM}
 
\begin{figure}[htbp]
\begin{center}
\includegraphics[scale=0.55,angle=-90]{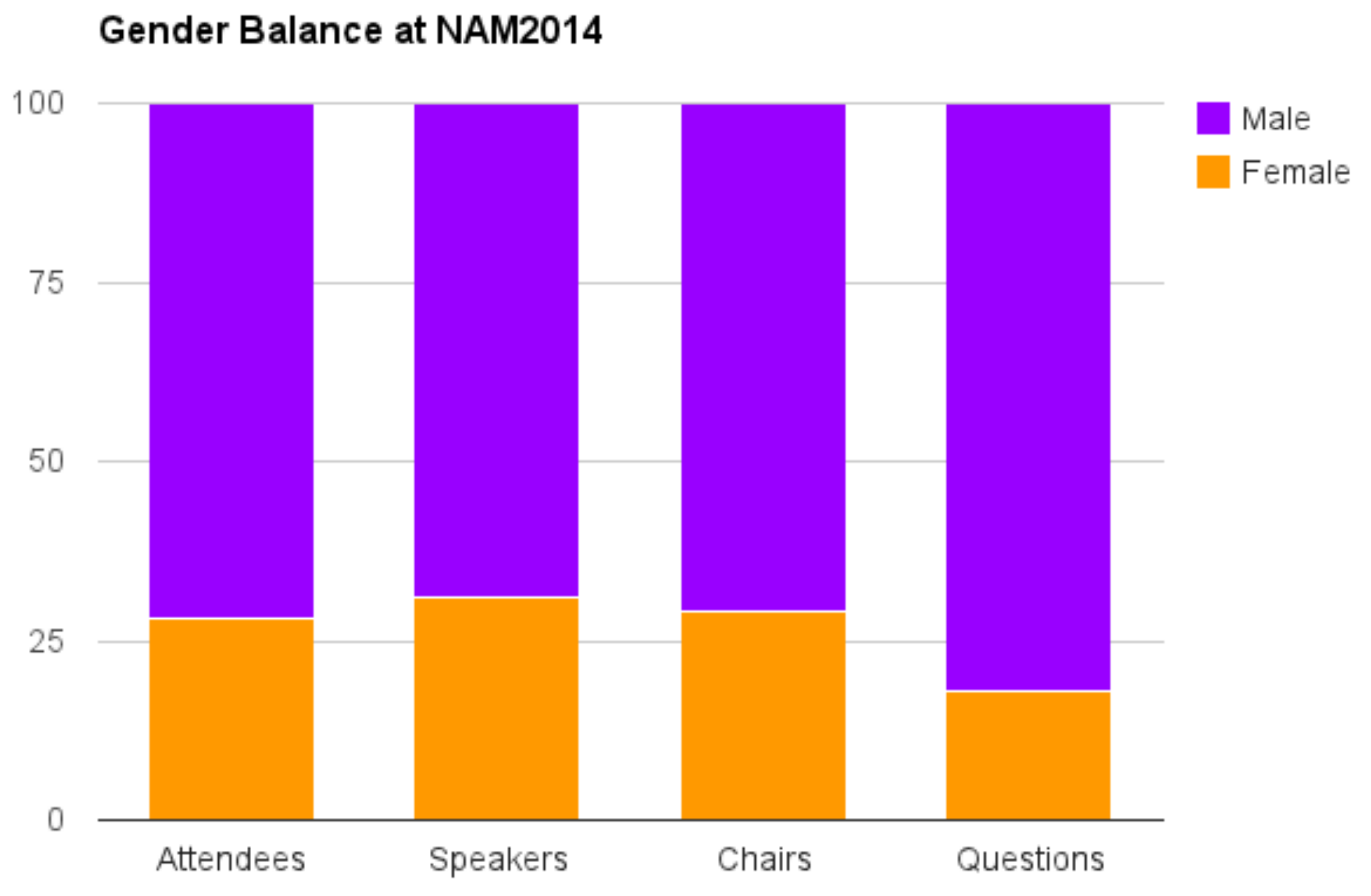}
\caption{Summary of the gender of attendees, speakers, chairs and question askers at NAM2014.}
\label{fig:summary}
\end{center}
\end{figure}

 We report that at NAM2014 the gender balance of speakers was 31$\pm$3\%\footnote{we report binomial uncertainties on these fractions} women and that of Chairs was 29$\pm$3\% women. This closely matched that of attendees, which we find to be made up of 28$\pm$2\% women. However women were observed to be  under-represented among question askers, which are made up of just 18$\pm$2\% women. These data are shown in Figure \ref{fig:summary}, and all our numeric data is summarised in Table \ref{data}.

\begin{table}
\caption{NAM2014 Data \label{data}}
\label{tab:summary}
\begin{tabular}{|l|c|c|c|} 
\hline
& Women & Men & Total  \\
\hline\hline
Attendees & 172 (28$\pm2$\%) & 452 (72$\pm2$\%) & 624 \\
Speakers$^*$ & 81 (31$\pm3$\%) & 181 (69$\pm3$\%) & 262 \\
Chairs$^*$ & 75 (29$\pm3$\%) & 188 (71$\pm3$\%) & 262\\
Question Askers (QAs)$^*$ & 101 (18$\pm2$\%) & 476 (82$\pm2$\%) & 577 (2.2$\pm$0.1 per talk)\\
~~~QA Female Speaker & 30 (17$\pm3$\%) & 150 (83$\pm3$\%)  & 180 (2.2 $\pm$0.1 per talk) \\
~~~QA Male Speaker & 71 (18$\pm2$\%) & 325 (82$\pm2$\%) & 396 (2.2 $\pm$0.1 per talk) \\
~~~QA Female Chair & 33  (22$\pm3$\%) & 119 (78$\pm3$\%) & 152 (2.1 $\pm$0.1 per talk) \\
~~~QA Male Chair & 68 (16$\pm2$\%) & 357 (84$\pm2$\%) & 425 (2.3 $\pm$0.1 per talk)\\
~~~Asking 1st question & 35 (14$\pm2$\%) & 216 (86$\pm2$\%) & 251\\
~~~Asking 2nd/3rd question & 49 (19$\pm3$\%) & 224 (81$\pm$3\%) & 273 \\
~~~Asking 4th/5th/6th/7th question &17 (32$\pm$6\%) & 36 (68$\pm$6\%) & 53 \\
\hline
\end{tabular}
\\$^*$These totals double count people who Chaired multiple talks (which is obviously common, given that most Chairs are for a session of several talks), gave multiple talks, and also those who asked multiple questions.
\end{table}

We define $p_F$ as the probability that a given question will be asked by a woman. Under the assumption of binomial statistics for the gender of a question asker, an estimate of this probability can be found from the fraction of questions asked by women (i.e. we estimate $p_F = 0.18\pm 0.02$). Obviously this value should equal the fraction of female attendees ($f_F=0.28\pm0.02$) if questions are equally likely to be asked by men and women, which is not found to be the case. 

From the data in Table \ref{tab:summary}, we see that the mean number of questions per man was $\langle M\rangle=1.05\pm0.03$ and per women was $\langle F\rangle=0.59\pm0.02$. If men and woman asked questions equally these numbers would be the same, but clearly they are not. The ratio of these numbers is $\langle M\rangle/\langle F\rangle=1.79\pm0.06$ -- a finding that male astronomers at NAM were roughly 1.8 times more likely to ask questions than female astronomers. Put another way, these data suggest that if NAM had had equal attendance by both men and women (i.e.\ if 50\% of the attendees had been women), and in that circumstance the likelihood of a woman asking a given question remained the same as we have measured here, then almost two-thirds of questions would still be asked by male astronomers. 

\begin{figure}[htbp]
\begin{center}
\includegraphics[scale=0.55,angle=-90]{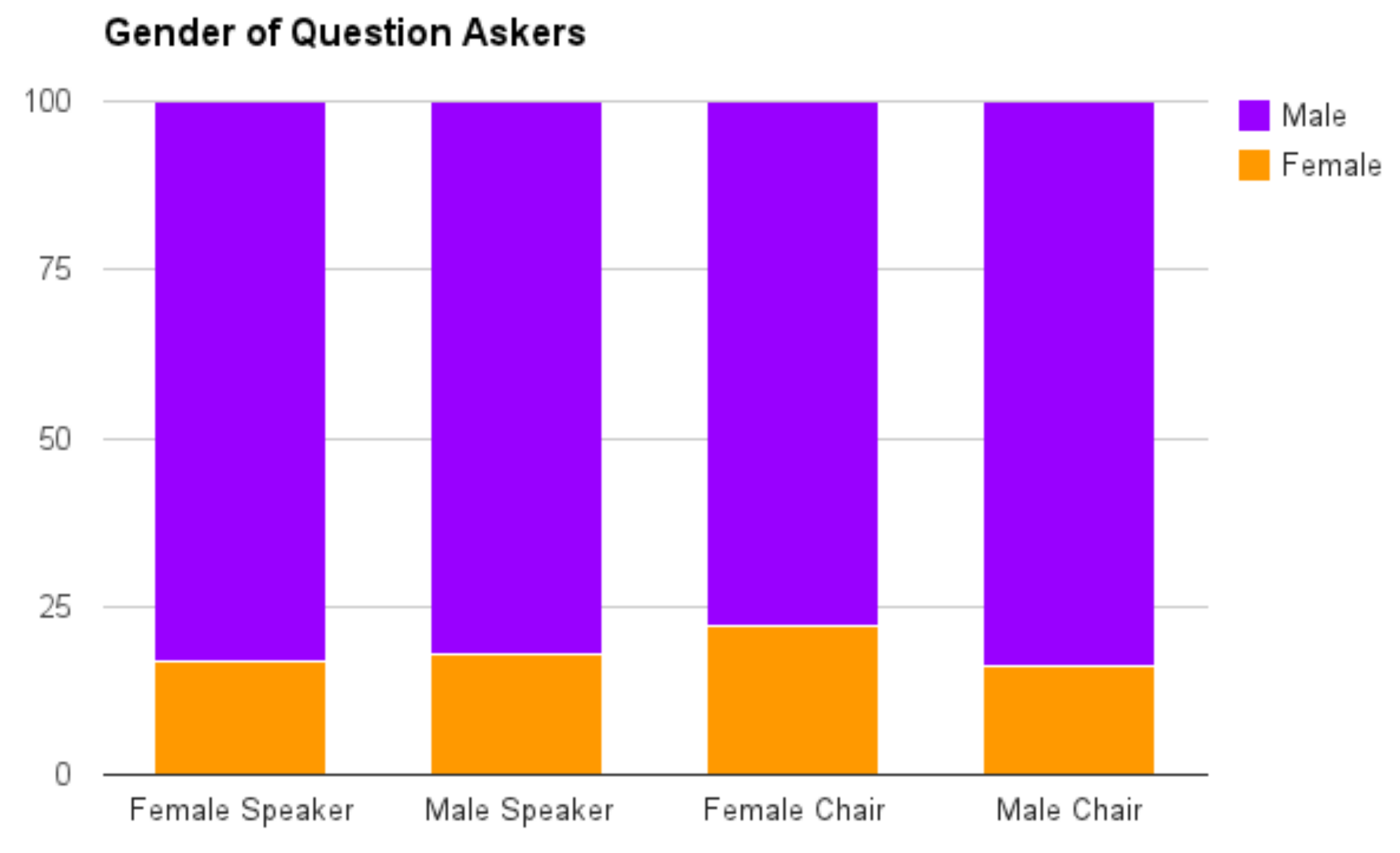}
\caption{Breakdown of gender of question askers by gender of speaker or chair.}
\label{fig:nam_gender2}
\end{center}
\end{figure}

We break this down further, looking separately at the impact of the gender of both the speaker and the chair on the gender balance of question askers. We find no difference in the gender balance of question askers separated by gender of speaker (with 17$\pm3$\% and 18$\pm2$\% questions by women for female and male speakers respectively), and a small difference detectable with the gender of the chair (22$\pm3$\% and 16$\pm2$\% questions were asked by women in talks Chaired by women or men respectively). These data are shown in Figure \ref{fig:nam_gender2}. This results in factors of $1.9\pm0.1$ and $1.7\pm0.1$ for how much more likely men are to ask questions than women of female and male speakers respectively; and factors of $1.4\pm0.1$ and $2.0\pm0.1$ for how much more likely men are to ask questions than women when there was a female or male Chair respectively.

\begin{figure}[htbp]
\begin{center}
\includegraphics[scale=0.38]{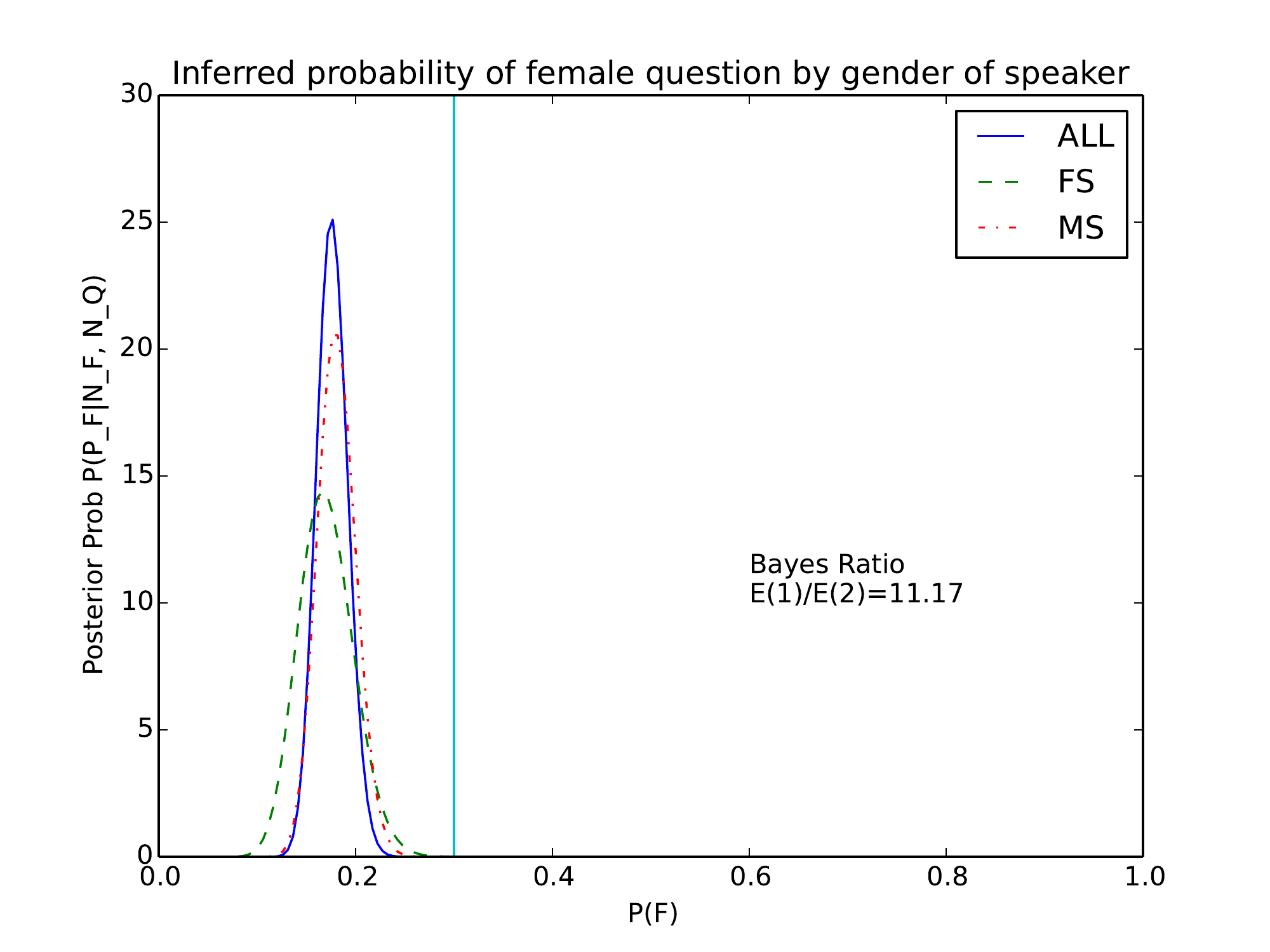}
\includegraphics[scale=0.38]{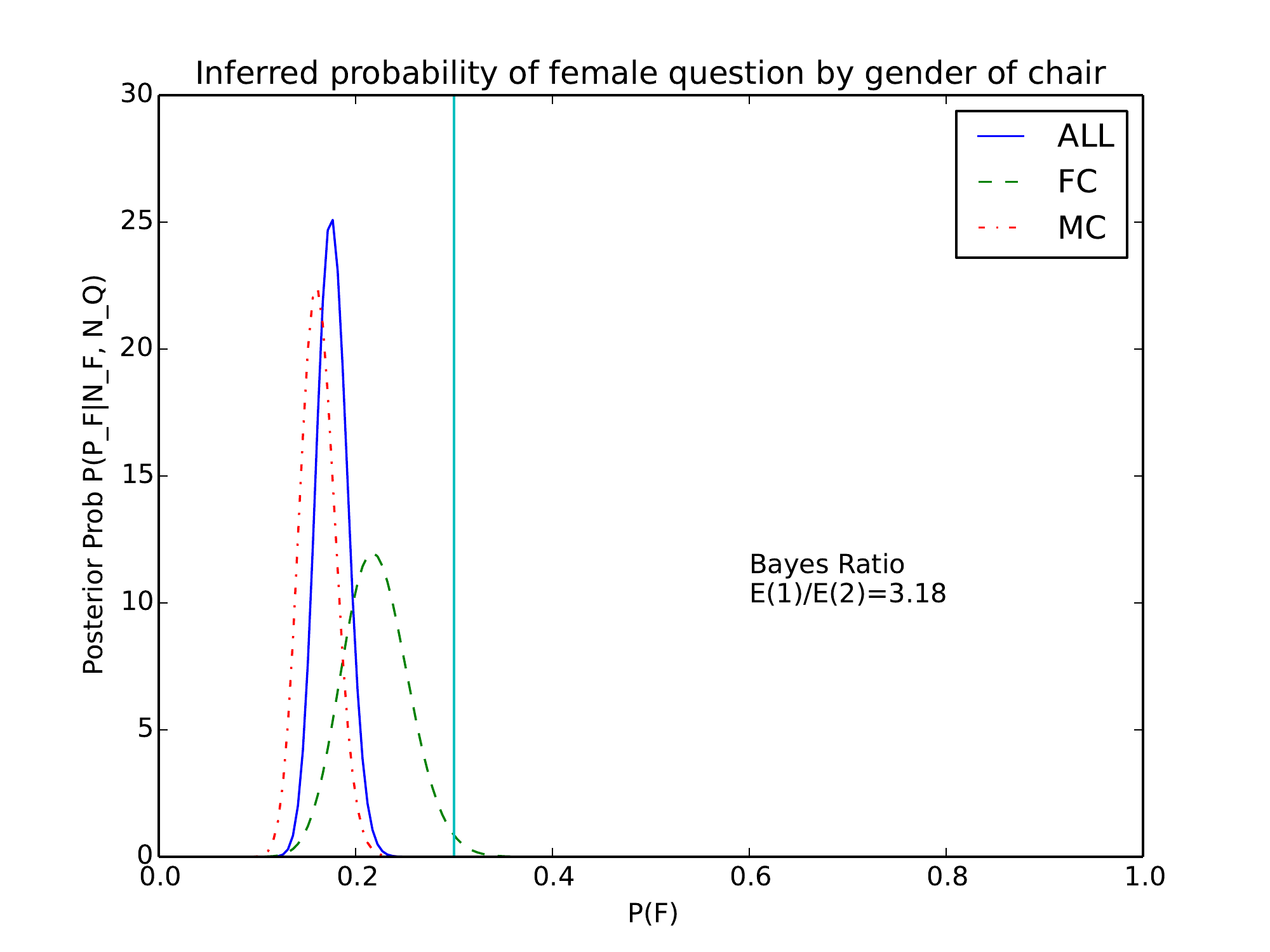}
\caption{Posterior probability of $p_F$, the probability that any individual questioner will be of female. We compare the data for the full conference with that of subsets split by male or female speaker (left) or chair (right). The vertical line indicates the fraction of female attendees at NAM2014. Also labelled is the Bayesian evidence ratio comparing our one and two parameter models.}
\label{fig:bayes1}
\end{center}
\end{figure}

We use Bayes theorem to assess the statistical significance of these data by comparing a single parameter model, in which the gender of the speaker or the gender of the chair has no impact on the probability $p_F$ that questions will be asked by women, to a two parameter model where there is a gender based effect. We find strong support (a Bayesian evidence or ``odds" ratio $R=11.2$) for a model in which the gender of the speaker has no impact, but we are not able to tell from these data if the gender of the chair has any impact (we find $R=3.2$, weakly favouring the ``no difference'' model). Figure \ref{fig:bayes1} shows the posterior probability distribution for each of these models.

\begin{figure}[htbp]
\begin{center}
\includegraphics[scale=0.38]{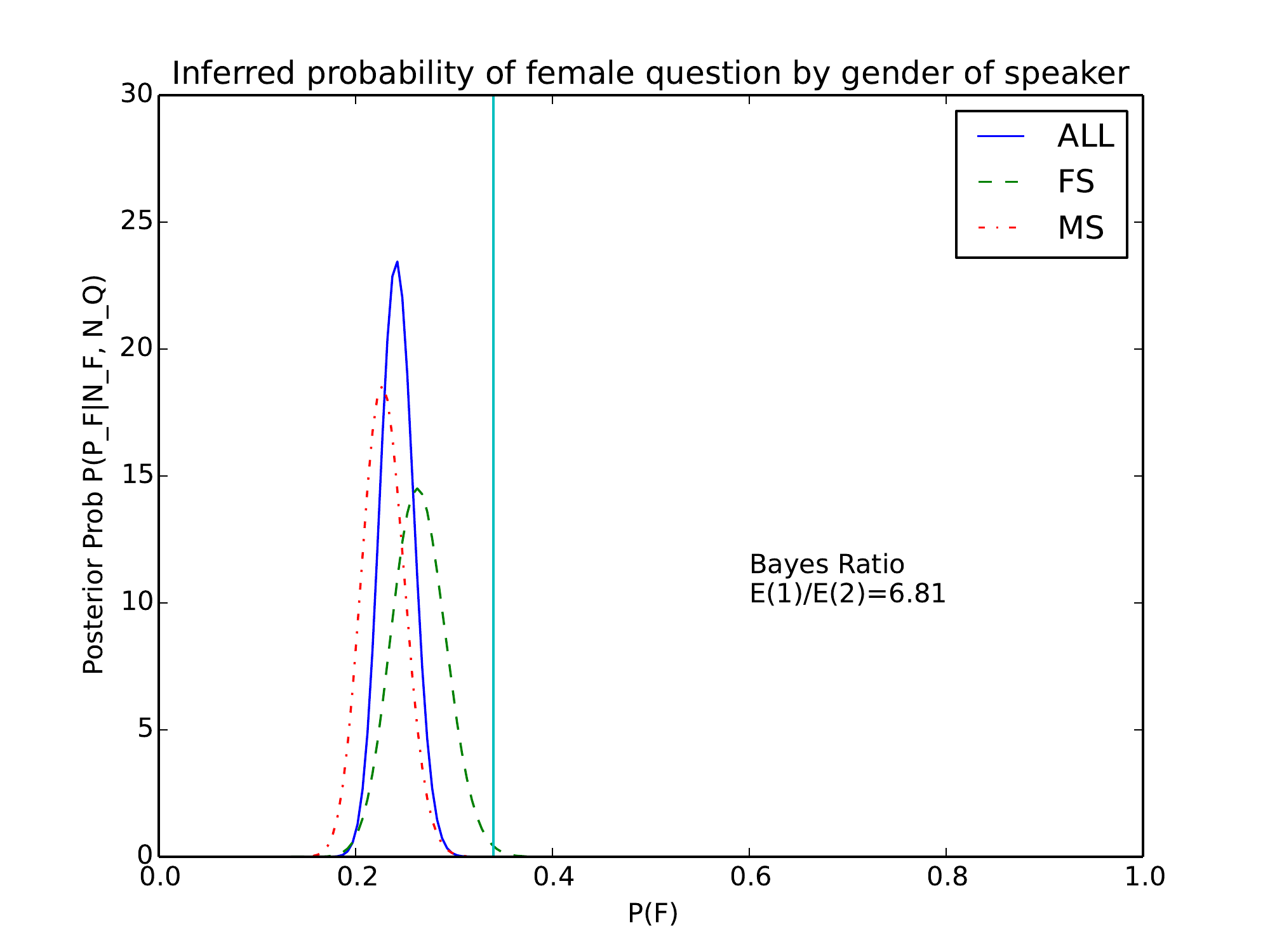}
\includegraphics[scale=0.38]{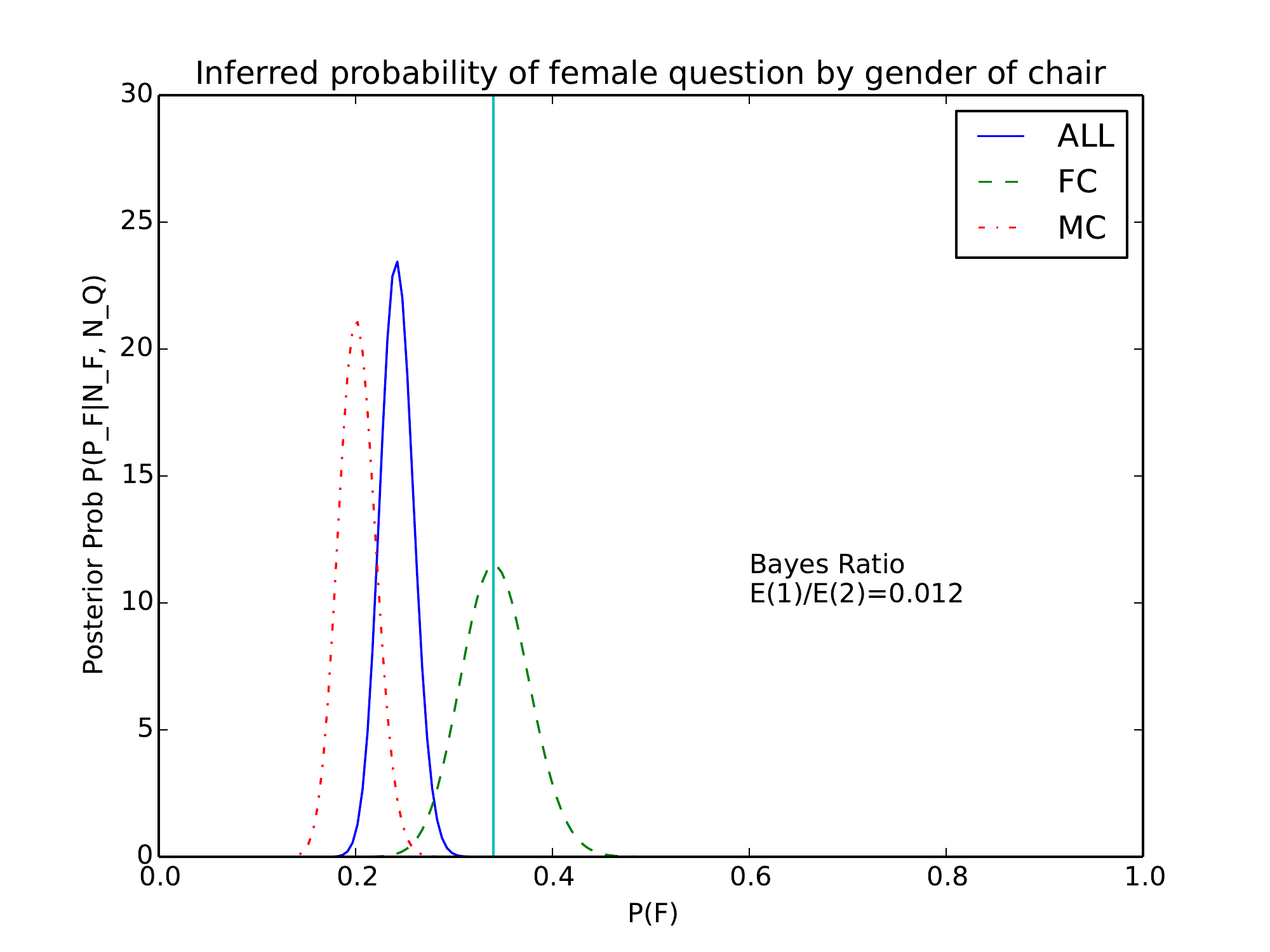}
\caption{Posterior probability of $p_F$, the probability that any individual questioner will be female for the AAS data. Curves are compared for the full conference and for a male or female speaker (left) or chair (right). The vertical line indicates the fraction of female attendees at AAS223. }
\label{fig:bayes2}
\end{center}
\end{figure}

This conclusion differs between our data and those collected by Davenport et al. (2014), so we run the same analysis on the data from AAS, as shown in Figure \ref{fig:bayes2}. Again we find good evidence that the gender of the speaker has little impact on the gender balance of the question askers ($R=7$), but now we favour a difference in the probability of women asking questions when the chair is female versus male ($R=1/0.012=83$ in favour of a difference). In fact, female chairs at the AAS were able to solicit questions from women and men with equal probability (i.e. $p_F\sim f_F$). 

\section{The Number of Questions asked at NAM} 

\begin{figure}[htbp]
\begin{center}
\includegraphics[scale=0.38]{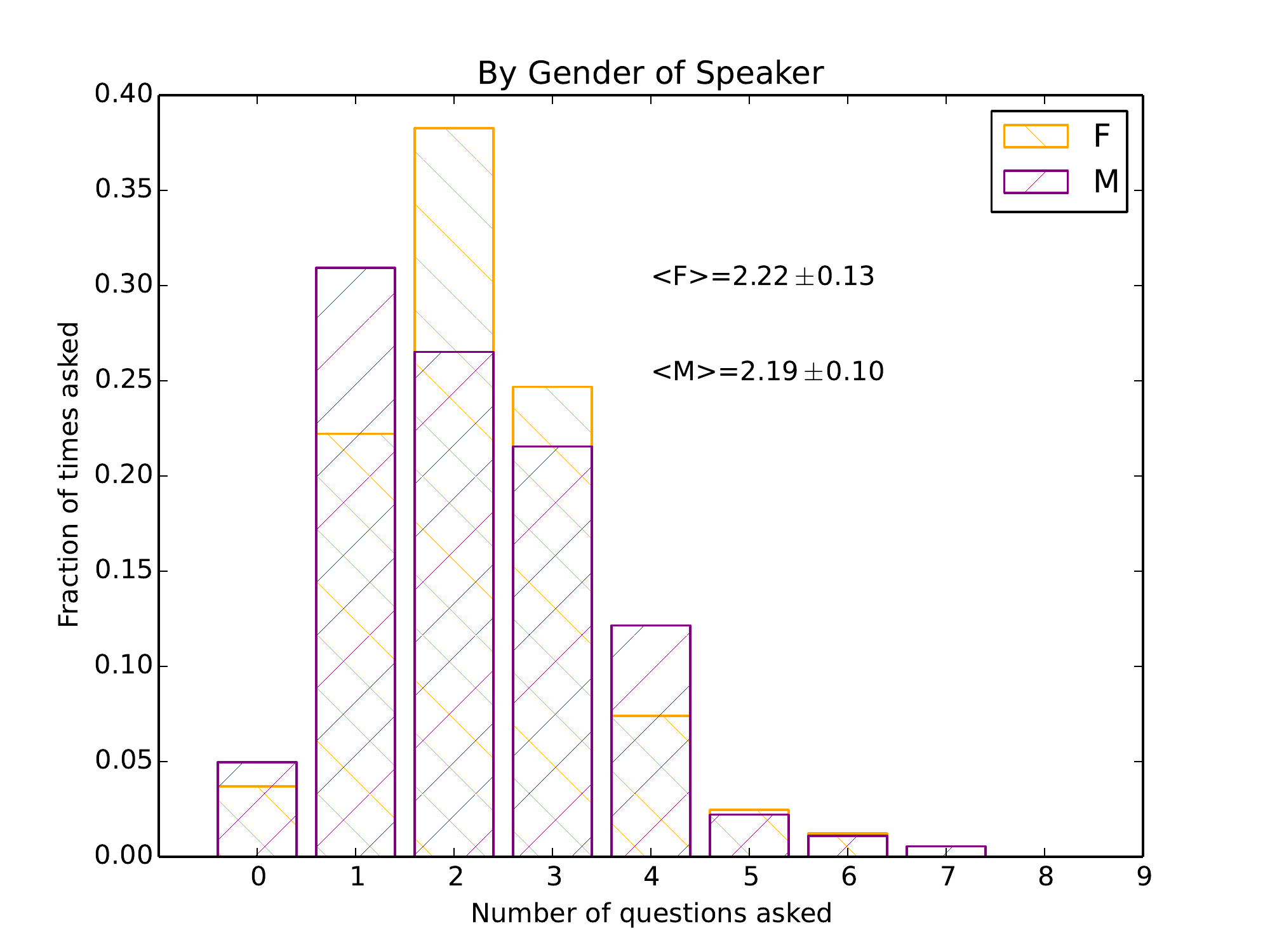}
\includegraphics[scale=0.38]{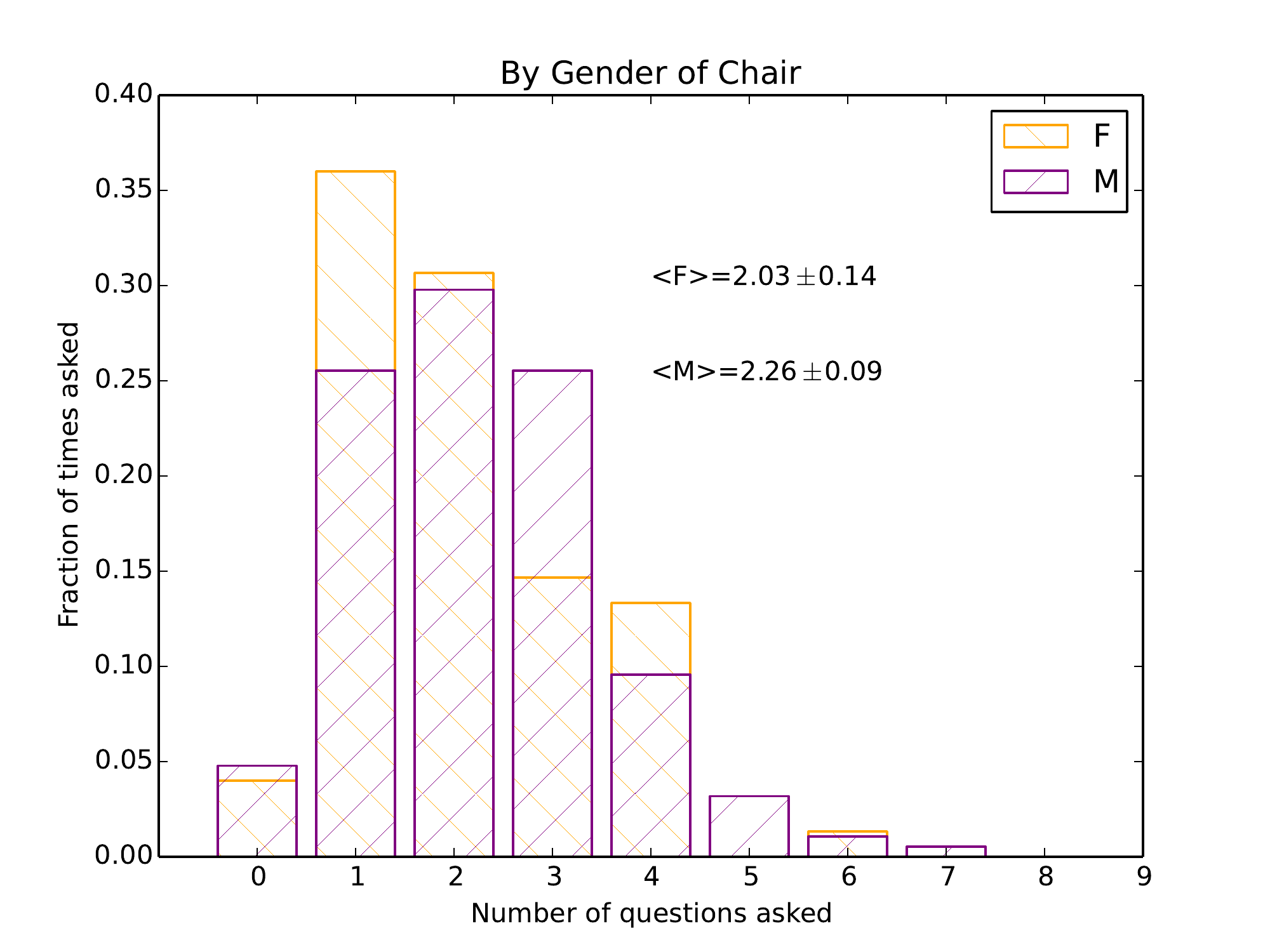}
\caption{Distribution of number of questions asked broken down by gender of speaker (left), chair (right). In each case, the distribution is normalised by the total number of questions at the conference from sessions with a speaker/chair of that gender. }
\label{fig:speaker_fig}
\end{center}
\end{figure}

On average we find that the number of questions asked at any given NAM talk was $N=2.2 \pm 0.1$. We find no detectable difference in the number of questions asked of female and male speakers (both $N=2.2\pm0.1$), but that female chairs solicited slightly smaller numbers of questions than male chairs ($N=2.0 \pm 0.1$ for female chairs, compared to $N=2.3 \pm 0.1$ for male chairs). The distributions of the numbers of questions asked are shown in Figure \ref{fig:speaker_fig}. Our result on this differs to AAS survey who found that female speakers were asked noticeably more questions ($N=3.3 \pm 0.2$) than their male counterparts ($N=2.6 \pm 0.1$). We note that these data also reveal that on average $0.7\pm0.2$ more questions are asked per AAS talk than per NAM talk ($N=2.9 \pm 0.1$ versus $N=2.2 \pm 0.1$). This difference might be caused by the different meeting formats (i.e. that talk times are very short at AAS - typically 5 mins per talk, and tend to be longer at NAM), or it could reveal cultural difference in US versus UK astronomers.

\subsection{Do Men and Women Ask Questions at the Same Time?}

\begin{figure}[htbp]
\begin{center}
\includegraphics[scale=0.55]{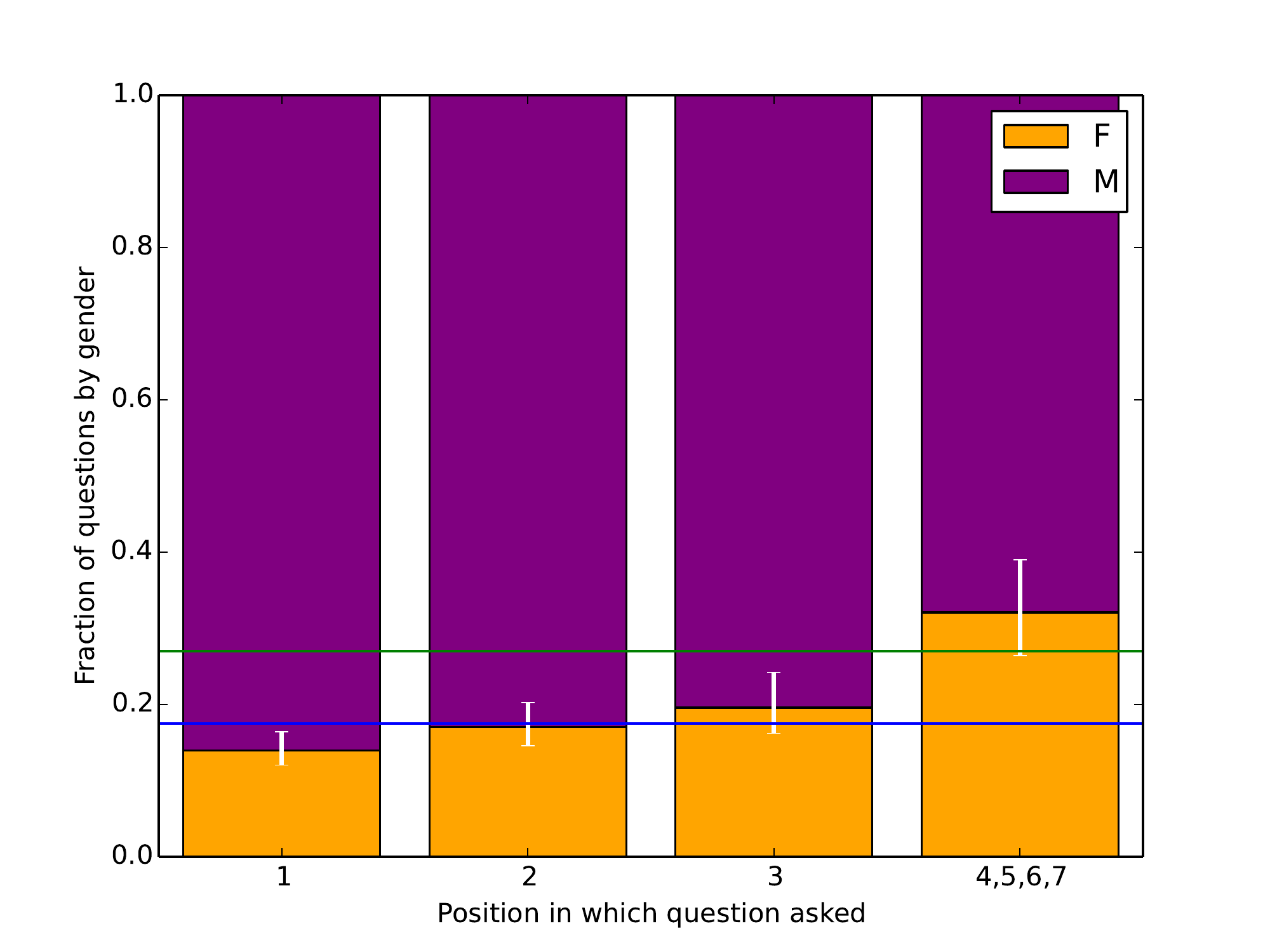}
\caption{Gender fraction of questions asked in the $n$th position following a talk. Error bars indicate the central 68\% confidence range based on inferring the gender ratio given the observed number of talks. The blue horizontal line at 0.18 indicates the gender ratio of all questions together, while the green line is at 0.28, indicating the gender ratio of all attendees at NAM.}
\label{fig:order_fig}
\end{center}
\end{figure}

The survey also allows us to test if men and women typically ask questions at the same time in a string of questions - we find that they do not. The NAM data show that female astronomers were more likely to ask questions later. The median question position for female question askers is $2.26\pm0.13$, while for men it is $1.89\pm0.05$. This difference is largely driven by men being much more likely to ask the first question after a talk. We show in Figure \ref{fig:order_fig} that male astronomers at NAM asked the first question 86\% of the time (or 6 times more often than women do), while by the time four questions or more are asked, women and men are equally represented as question askers  -- 32\% of the 4th/5th/6th/7th questions (aggregated due to low numbers) are asked by women. 

\section{Discussion}

\subsection{Comparison to AAS Results}

The UK and US astronomical communities have many similarities, but are not the same. As George Bernard Shaw, and later Winston Churchill, said, we are ``two nations divided by a common language." In the most recent RAS demographic survey \citep{mcwhinnie2011}, the UK Astronomy community comprised a total of almost 1700 people with the proportion of women varying from almost 30\% among postdocs to just 7\% of full Professors in Astronomy. The US astronomical community is somewhat larger - the most recent demographic survey of the AAS \citep{anderson2013} lists 2523 members of whom 25\% are female. Internationally, the UK and US astronomical communities (as tracked by IAU membership; \citealt{cesarksy2010}) are seen as having very similar gender balance, with 11.6\% and 12.1\% IAU members being women respectively (this relatively low fraction of women among IAU members is usually attributed to the relative seniority of IAU member astronomers). 

We find subtle differences revealed in the question asking tendencies at the national conferences of the US (AAS223) and the UK (NAM2014). The US community ask slightly more questions on average ($0.7\pm0.2$ more questions per talk), and reveal a tendency to ask more questions of female speakers than male which is not observed in the UK community. In both communities we find that women are less likely than men to ask questions, and that when a session is chaired by a woman this can be improved (either slightly as in the UK community, or quite substantially as seen in the US data). 

The January AAS is a very different meeting to the annual NAM. It is much larger, and has many more parallel sessions in general. Contributed talks (with the exception of dissertation talks allowed once per career) are limited to 5 mins and sorted into groups by topic after the abstracts are submitted. The chairs are assigned to parallel sessions by the main organisers, and are required to attend training at the beginning of the meeting which includes advice on stimulating and moderating discussion following talks. At NAM, by contrast, parallel session topics are proposed to the SOC in advance of abstract submission, which must then be submitted to a specific topic. Once a session is accepted, the proposers (who are typically, but not always, the Chair of their session) determine the length of talks, which can therefore vary substantially from session to session. At NAM2014, the Local Organising Committee emailed session chairs with extensive informal advice on promoting inclusive discussion.  All of these differences can change the culture and environment of a meeting in subtle ways, on top of any cultural differences between the US and UK.

\subsection{Implications for UK astronomy} 

 The situation for female astronomers in the United Kingdom has come a long way since 1835 when the Royal Astronomical Society admitted its first women as Honorary Fellows (Caroline Herschel and Mary Somerville; full fellowship was open to women from 1915). Instances of overt discrimination are now thankfully rare in our community (and have been illegal since 1975), but subtle issues like unconscious bias (demonstrated to affect men and women equally, e.g. \citealt{Steinpreis}), as well as stereotype threat \citep{Spencer,Betz} and the sheer fact of the raw numbers can still make the astronomical community feel like a hostile place for women, despite the best intentions of all involved.  
  
 It is a fact that less than a third of professional astronomers in the UK are women \citep{mcwhinnie2011}, and this drops to fewer than one in ten among the most senior (full) Professors. There is a rule-of-thumb,  that members of a ``minority group" will stop noticing they are in a minority once more than about a third of people they interact with is made up of people from the group they identify with. The UK astronomical community may be about to reach this tipping point (and the US community is just past it), so it will be fascinating to see if any changes occur. 
  
\begin{figure}[htbp]
\begin{center}
\includegraphics[scale=0.6,angle=-90]{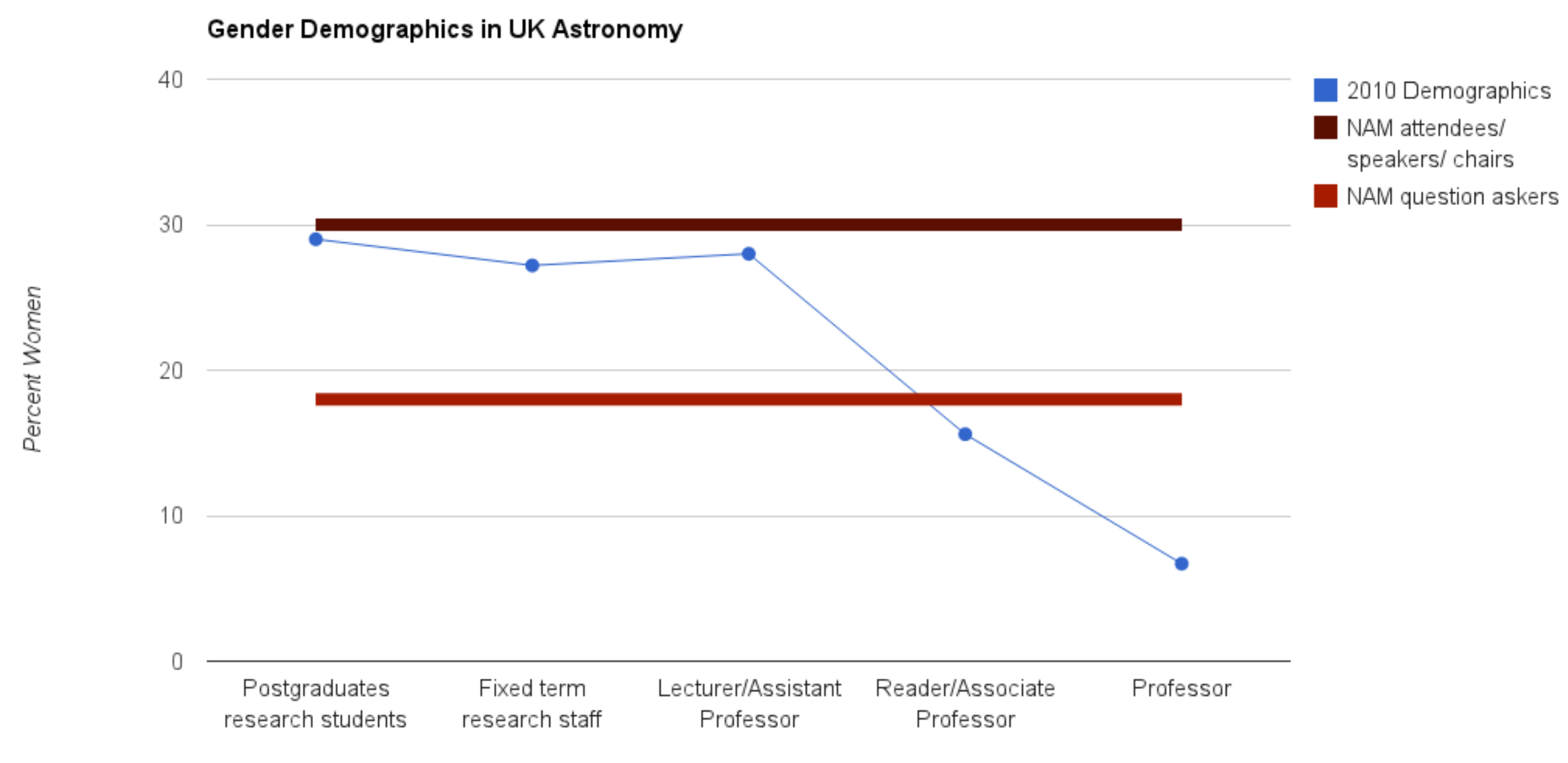}
\caption{The gender balance of NAM attendees, speakers, chairs and question askers compared to the gender balance by professional grade for the UK astronomy community.}
\label{fig:demographics}
\end{center}
\end{figure}

  We have found that at the UK National Astronomy Meeting held in 2014 women were around 1.8 times less likely then men to ask questions. A similar observation at the AAS was interpreted as likely being due to question askers being more senior than speakers and attendees in general at the conference (and the lower fraction of women among more senior astronomers).  We show in Figure \ref{fig:demographics}, that the gender balance of NAM attendees, speakers and chairs roughly matches that found in the UK astronomy community in postgraduate research students, postdocs and most junior academic staff (i.e.\ Lecturers or Assistant Professors in the US terminology) as reported in the 2010 RAS Demographic Survey (McWhinnie 2011), while the gender balance of those asking questions at NAM matches the gender balance found amongst UK-based Readers (or Associate Professors). As a community we should be seeking to make this event a welcoming, diverse and equal opportunity for intellectual discourse about our field regardless of gender or academic status. Our results suggest that either women, and/or more junior people attending NAM appear to feel less able to ask questions than their more senior and/or male colleagues which we do not believe is desirable. 
   
\subsection{The Psychology of Asking Questions}

 Studies of the psychology of asking questions tend to focus on student participation in classroom discussions (e.g. Krupnick 1985, Younger et al. 1999 or see \citealt{murphy2006} for physics specific classrooms), or the contributions to conversation in general (e.g. see the review by \citealt{james1993}) rather than question-asking at professional research conferences. Never-the-less it appears that a lower engagement from women than men is a fairly ubiquitous finding in such studies which tend to conclude that men/boys on average dominate most kinds of mixed discussions \cite{krupnick1985,james1993,younger1999}, and if women participate equally they risk being perceived (negatively) as dominating the conversation.  We consider here if any general conclusions or ideas for good practice can be drawn from this rich social science literature on gender and discourse. 
  
  There is much discussion of creating an environment conducive to participation so that people feel ``psychologically safe in taking the risk of asking a question".  Research on the impact of stereotype threat, or the risk of being judged by negative stereotypes of the ability of women in maths and science \citep[e.g.][]{Spencer} do support the idea that female astronomers (on average) would tend to find it harder to ask questions, for example \citep{carr2010stereotype} found that women facing stereotype threat are more risk averse than typical. In addition, there are suggestions that subtle differences in criticism from teachers over time can lead girls/women to develop low self confidence in their abilities, while boys/men may (again on average) overestimate theirs (e.g. \citealt{dweck1986}). This boost in self confidence given to male astronomers over the course of their schooling may make them more likely to be willing to ask questions. Female astronomers (on average) may simply need more encouragement and help to ask questions. 
  
 \citet{sunderland2000new} cautions against the representation of women as ``victims" of male dominance in discourse, for example pointing out there is a difference between quantity and quality of interactions, and reporting in her work (based on observations of language learning classrooms) that it was a just a small fraction of the boys who were dominating the discourse. Judging how good questions are following a talk would be difficult, and risks subjective assessments (prone to unconscious bias). We suggest the length of response from the speaker could be taken as an imperfect but objective measure.  It would also be difficult to record who is asking the questions, unless question askers were encouraged to identify themselves by name at the start of their question.
 
 Several results have found that the presence of a female role model has a positive impact on female participation (e.g. Krupnick 1985). In our NAM survey, the impact of a female Chair had only a mild impact on the gender balance of question askers, but in the AAS survey this was found to make a very significant difference. Based on these results, ensuring a good gender balance of session Chairs can only help improve the environment to encourage women to ask questions. Favouring women to ask the first question may also have a simple positive impact.
 
  In classroom interactions positive effects on gender equal participation have been shown when teachers wait longer to get answers, and don't immediately go to the first person who raises their hands \citep{murphy2006}. It's curious that in the NAM data we show women are as likely as men to ask later questions. Providing session Chairs with strategies to maintain discussion and encourage longer Q\&A sessions may have a positive impact on the participation of female astronomers in asking questions.

\subsection{Suggestions for Action}

The findings of this first survey of the gender balance of question asking at NAM may have raised more questions than it answered. While we detect a clear gender difference in attendees and those asking questions, given what we know about how the gender balance changes with age/seniority in UK astronomy we are unable to tell if this is caused directly by gender differences or by age/seniority differences in question asking behaviour. If this survey is replicated in future it would help to have access to both the gender balance and the seniority of attendees of NAM. We found that women were more likely to ask later questions in a string, but did not record how often Q\&A sessions were cut short before all questions were asked and if this behaviour had any impact on the overall gender balance. We also did not record if the Chairs routinely ask questions, and if this behaviour shows any gender differences. 

We propose the following set of actions based on the findings of this work and our review of the literature on the psychology of question asking: 

\begin{itemize}
\item A similar survey should be repeated at future NAMs (perhaps not every year, but at least within the next few years), and monitored for changes. In that way we can test if our suggested actions have any impact, as well as disentangle the roles of gender and seniority in asking questions. In future surveys we recommend that the following additional data be collected: 
\begin{itemize}
\item A genuinely unique identifier for each talk (preferably assigned beforehand to enable a drop down menu in the form). 
\item If the Chair asked a question, and when (i.e. collect a string such as `FFCM' where C represents a question asked by the Chair).
\item Was the Q\&A session cut short, or did it end due to lack of further questions? 
\item If possible, to record the name of the question asker as well as their gender. 
\item The gender and seniority of attendees at NAM. 
\end{itemize}
\item Chairs of Sessions should be given a brief training session or sent guidelines with advice for good practice, which we suggest should include the following recommendations: 
\begin{itemize}
\item Younger scientists should be explicitly encouraged to ask questions (i.e. this should be stated in introductory remarks by the Chair), and favoured if there is a choice of question askers. 
\item If there is a choice between male and female question askers for the first question, a question from a woman should be given priority. 
\item Question askers should be asked to identify themselves by name. 
\item If possible, Q\&A sessions should not be cut short before at least 4 questions have been asked (if they need to be ended early). To enable this session organisers should schedule enough time for questions and speakers should not be allowed to run over time. 
\end{itemize}
\end{itemize} 
We believe these actions will help to make our annual meeting a more open opportunity for discourse among professional astronomers regardless of their gender or seniority.


\section*{Acknowledgements} 
\label{sec:ack}

{\it Jonathan Pritchard, Imperial Centre for Cosmology Inference, Imperial University; j.pritchard@imperial.ac.uk. 
Karen Masters, Institute for Cosmology and Gravitation, University of Portsmouth; karen.masters@port.ac.uk.
James Allen, Sydney Institute for Astronomy (SIfA), School of Physics, The University of Sydney; j.allen@physics.usyd.edu.au.
Filippo Contenta, Department of Physics, University of Surrey; f.contenta@surrey.ac.uk.
Leo Huckvale, Jodrell Bank Centre for Astrophysics, University of Manchester; leo.huckvale@postgrad.manchester.ac.uk.
Stephen Wilkins, Astronomy Centre, Department of Physics and Astronomy, University of Sussex; s.wilkins@sussex.ac.uk.
Alice Zocchi, Department of Physics, University of Surrey; a.zocchi@surrey.ac.uk.

We'd especially like to thank Bob Nichol, the NAM LOC and the RAS Diversity Committee for allowing us to conduct this survey at NAM 2014 in Portsmouth. The NAM Hack day organised by Robert Simpson from the Zooniverse/Oxford University and Arfon Smith from Github provided us with the opportunity to share ideas and get the project off the ground. It also inspired the use of GitHub for organising our sharing of data and progress. Special thanks go to James Davenport for sharing the web form used in the AAS 223 survey. Finally, we thank all those NAM attendees that submitted data for this survey without whom this would not have been possible.}

 \bibliography{nambib}

 
 \end{document}